\documentclass[aps,prb,twocolumn,superscriptaddress,showpacs]{revtex4}

\usepackage[dvips]{graphicx,color}
\usepackage{dcolumn}
\usepackage{bm}

\begin{document}

\title{
Change in lattice modulation upon gigantic magnetoelectric transition in GdMnO$_3$ and TbMnO$_3$
}

\author{T. Arima}
\affiliation{Institute of Multidisciplinary Research for Advanced Materials,
Tohoku University, Sendai 980-8577, Japan}
\affiliation{
Spin Superstructure Project, ERATO, Japan Science and Technology Agency, 
AIST Tsukuba Central 4, Tsukuba 305-8562, Japan
} 

\author{T. Goto}
\affiliation{
Department of Applied Physics, University of Tokyo,
Tokyo 113-8656, Japan
}
\author{Y. Yamasaki}
\affiliation{
Department of Applied Physics, University of Tokyo,
Tokyo 113-8656, Japan
}
\author{S. Miyasaka}
\affiliation{
Department of Applied Physics, University of Tokyo,
Tokyo 113-8656, Japan
}
\author{K. Ishii}
\affiliation{
Synchrotron Radiation Research Center, Japan Atomic Energy Research
Institute, Hyogo 679-5148, Japan
}
\author{M. Tsubota}
\affiliation{
Synchrotron Radiation Research Center, Japan Atomic Energy Research
Institute, Hyogo 679-5148, Japan
}
\author{T. Inami}
\affiliation{
Synchrotron Radiation Research Center, Japan Atomic Energy Research
Institute, Hyogo 679-5148, Japan
}
\author{Y. Murakami}
\affiliation{
Synchrotron Radiation Research Center, Japan Atomic Energy Research
Institute, Hyogo 679-5148, Japan
}
\affiliation{
Department of Physics, Tohoku University, Sendai 980-8578, Japan
}
\author{Y. Tokura}
\affiliation{
Spin Superstructure Project, ERATO, Japan Science and Technology Agency, 
AIST Tsukuba Central 4, Tsukuba 305-8562, Japan
}
\affiliation{
Department of Applied Physics, University of Tokyo,
Tokyo 113-8656, Japan
}



\begin{abstract}
Single-crystal synchrotron x-ray diffraction measurements in strong magnetic fields have been performed for magnetoelectric compounds GdMnO$_3$ and TbMnO$_3$.  
It has been found that the $\makebox{\boldmath $P$} \parallel a$ ferroelectric phase induced by the application of a magnetic field at low temperatures is characterized by commensurate lattice modulation along the orthorhombic $b$ axis with $q=1/2$ and $q=1/4$.  
The lattice modulation is ascribed to antiferromagnetic spin alignment with a modulation vector of (0 1/4 1).  
The change of the spin structure is directly correlated with the magnetic-field-induced electric phase transition, because any commensurate spin modulation with (0 1/4 1) should break glide planes normal to the $a$ axis of the distorted perovskite with the $Pbnm$ space group.  
\end{abstract}


\maketitle


Interest has been recently enhanced in multiferroic materials, where magnetic and electric ordering can coexist.\cite{Fiebig}
Among such materials, rare-earth manganites $R$MnO$_3$  ($R=$Gd, Tb, and Dy) with orthorhombically distorted perovskite structure are prototypical and show unusual interplay of magnetic and electric properties.\cite{Kimura_nature,Goto,Popov,Noda,Kimura_PRB2}    
The compounds show gigantic magnetoelectric (ME) effects:  
Electric polarization {\boldmath $P$} can be flopped by applying a magnetic field in Tb and Dy derivatives. GdMnO$_3$, whose orthorhombic distortion is less than TbMnO$_3$ and DyMnO$_3$, shows a magnetic-field-induced ferroelectric transition.  Giant magnetocapacitance was also realized in the vicinity of these phase boundaries.  
These ME effects are different from the conventional linear ME effect, which can be described with perturbation theories.  Therefore, it is important to understand the novel mechanism of the ME phenomenon in a series of perovskite-type $R$MnO$_3$ compounds.  
One of the most important clues to solve this problem is the nature of change in spin structure with magnetic fields.  
It has been reported that the ferroelectric transition of TbMnO$_3$ and DyMnO$_3$ is accompanied by locking of the modulation vector of the antiferromagnetically ordered spin system.\cite{Kimura_PRB1,Kimura_nature,Goto,Kajimoto}
This result strongly suggests that the spin structure correlates with the ferroelectricity in orthorhombic $R$MnO$_3$ compounds.  
This paper reports a synchrotron x-ray diffraction study of GdMnO$_3$ and TbMnO$_3$ in strong magnetic fields.  Synchrotron x-ray diffraction is one of the most powerful tools to investigate this class of magnetoelectrics, because it can probe displacements of oxygen ions by the order of 0.001 \AA, expected from the electric polarization as large as 500--1000 $\mu$C/m$^2$.  
We have found a systematic change in the modulation vector of the lattice, which should be connected to the antiferromagnetic modulation vector.  

Single crystals of $R$MnO$_3$ with $R=$Tb and Gd were grown by a floating zone method.\cite{Kimura_PRB1}   The obtained crystal boules were cut into thin plates with the widest faces of (0 1 0).  (Miller indices are based on a $Pbnm$ space group in this paper.)
Nonresonant single-crystal x-ray diffraction was measured at the Beamline 22XU, SPring-8, JAPAN.  
X-rays from an undulator source were monochromated to 18 keV with a Si (111) double crystal and focused on the sample in a He-flow cryostat with a superconducting magnet, which was mounted on a two-axis goniometer.  
Superlattice reflections along the (0 $k$ 1) line in the reciprocal space were surveyed in a magnetic field $H_b$ applied along the $b$ axis.  The possible heating by x-ray irradiation was carefully checked by changing the incident beam intensity.

\begin{figure}
\includegraphics*[width=8.6cm]{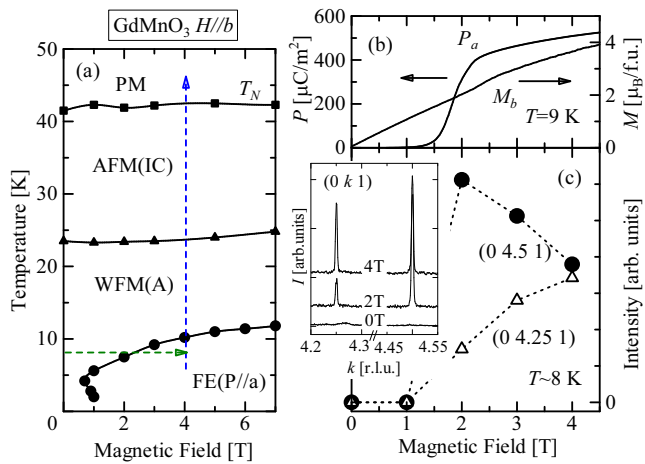}
\caption{
(Color online)
(a) Magnetic and electric phase diagram of GdMnO$_3$ with applied magnetic fields along the $b$ axis.  Squares, triangles, and circles indicate paramagnetic(PM)-to-incommensurate antiferromagnetic (AFM(IC)), AFM(IC)-to-canted A-type antiferromagnetic (WFM(A)), and  WFM(A)-to-{\boldmath $P$}$\parallel a$ ferroelectric (FE($P//a$)) phase transitions, respectively.  
X-ray studies were performed in the condition along broken lines.  
(b) Polarization along the $a$ axis ($P_a$) and magnetization ($M_b$) plotted against magnetic field applied parallel to the $b$ axis.  
(c) Intensities of superlattice reflections as a function of magnetic field.  Inset shows x-ray diffraction profiles along (0 $k$ 1) in some magnetic fields.  
}
\label{gmo}
\end{figure}


Figure \ref{gmo}(a) shows the $H_b$-$T$-plane phase diagram in GdMnO$_3$.  With increasing $H_b$, ferroelectric polarization parallel to the $a$ axis is induced as shown in Fig.~\ref{gmo}(b).   
The present x-ray study indicates that the ferroelectric phase is accompanied by a commensurate lattice modulation.  (See the inset of Fig.~\ref{gmo}(c).)   
The lattice modulation vector {\boldmath $q_\ell$} can be determined as $\makebox{\boldmath $q_\ell$}=0.25\makebox{\boldmath $b^*$}$ but not $0.75\makebox{\boldmath $b^*$}$ because the (051) reflection is forbidden in the $Pbnm$ space group.  

Lattice modulation in a series of $R$MnO$_3$ compounds has been ascribed to the exchange striction.\cite{Kimura_PRB1,Kimura_nature}  Antiferromagnetic arrangement of Mn spin moments with a modulation vector of (0 $q_{\rm Mn}$ 1) originates from spin frustration, and causes the (0 $2q_{\rm Mn}$ 0) lattice modulation.  The observed commensurate superlattice reflections in the present study should be also correlated with a spin modulation.  
The magnetic modulation wavenumber $q_{\rm Mn}$ in the $\makebox{\boldmath $P$}\parallel a$ phase is either 1/4 or 1/8.  
To determine the $q_{\rm Mn}$ value, intensities of both superlattice reflections are plotted in Fig.~\ref{gmo}(c).  
The relative intensity of the (0 4+1/4 1) reflection to (0 4+1/2 1) becomes large with increasing $H_b$.  
The lattice modulation expected from exchange striction can appear not only at $k=2q_{\rm Mn}$ but also at $k=q_{\rm Mn}$ in a high field, because a composite symmetry operation of translation of half of the magnetic unit cell and time reversal vanishes.  
As the magnetization and resultant symmetry breaking become larger, the $k=q_{\rm Mn}$ series of reflections should increase in intensity.  
We therefore conclude that the magnetic modulation vector in the ferroelectric phase of GdMnO$_3$ should be (0 1/4 1).

\begin{figure}
\includegraphics*[width=7cm]{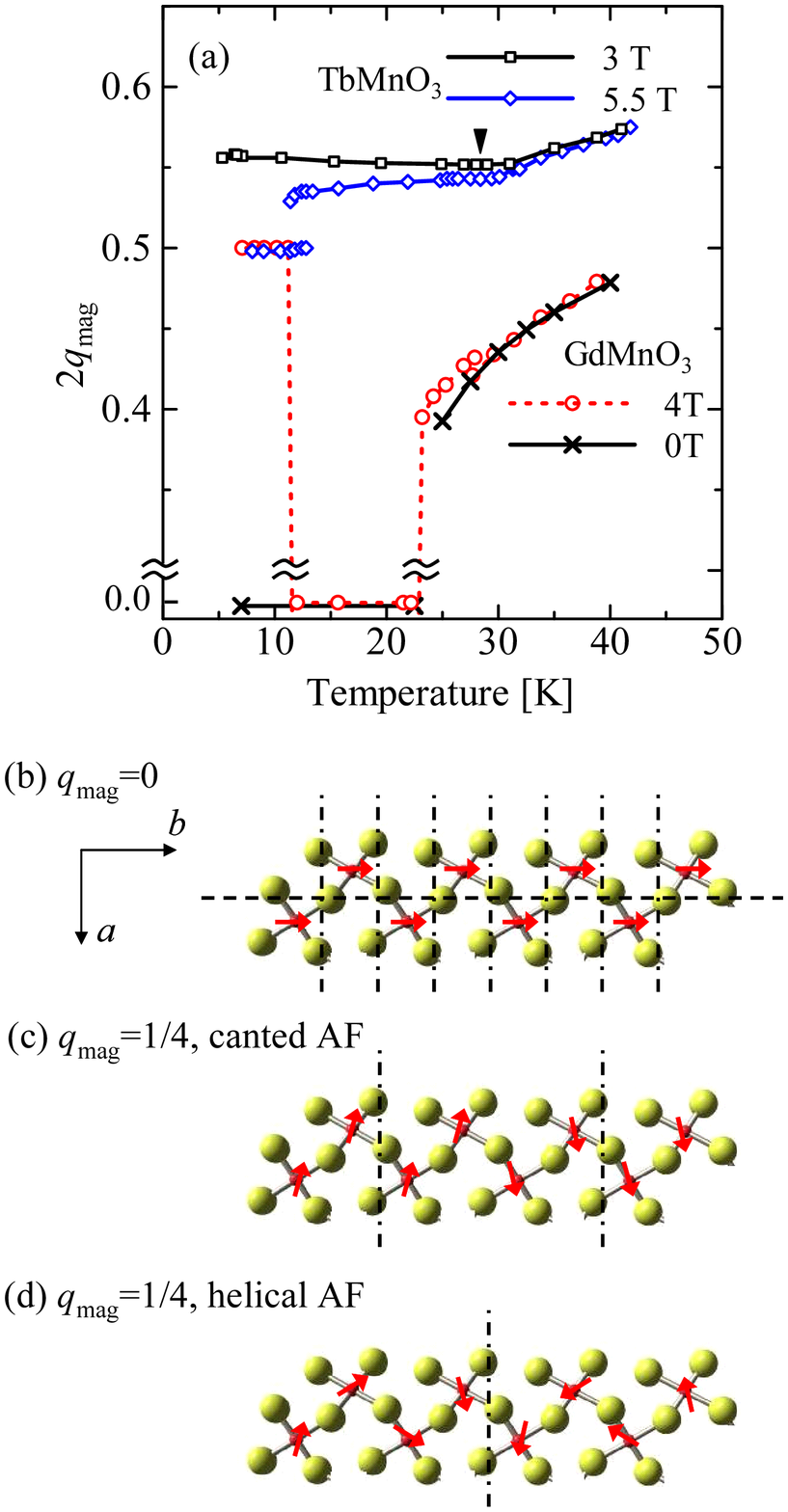}
\caption{
(Color online)
(a) Change in spin modulation vectors (0 $q_{\rm Mn}$ 1) of GdMnO$_3$ (TbMnO$_3$) with temperature in a magnetic field of 0 T and 4 T (3 T and 5.5 T).  A solid triangle indicates the ferroelectric Curie temperature of TbMnO$_3$.  
(b) Spin arrangement for A-type antiferromagnetic paraelectric phase in GdMnO$_3$. 
A broken line and dotted-broken lines show a $b'$-glide plane and $n'$-glide planes, respectively.
(c) Canted antiferromagnetic structure with a modulation vector of (0 1/4 1).  
Dotted-broken lines indicate $n$-glide planes.  
MnO$_2$ planes with this spin ordering stack along the $c$ axis with a shift of $2b$.  
(d) Helical antiferromagnetic structure with a modulation vector of (0 1/4 1).
A dotted-broken line indicates an $n$-glide plane.  
}
\label{model}
\end{figure}

Figure \ref{model}(a) shows temperature dependence of magnetic modulation vector in GdMnO$_3$ and TbMnO$_3$.  
With decreasing temperature in a magnetic field of 4 T, GdMnO$_3$ undergoes successive phase transitions.  
The $q_{\rm Mn}$ value of GdMnO$_3$ is not much dependent on $H_b$ in the paraelectric phase ($>12$ K).  
The appearance of commensurate lattice and spin modulation in the magnetic field below 12 K is in good accordance with the onset of the ferroelectric transition under the same condition (see the vertical dashed arrow in Fig.~\ref{gmo}(a)).  

\begin{figure}
\includegraphics*[width=8.6cm]{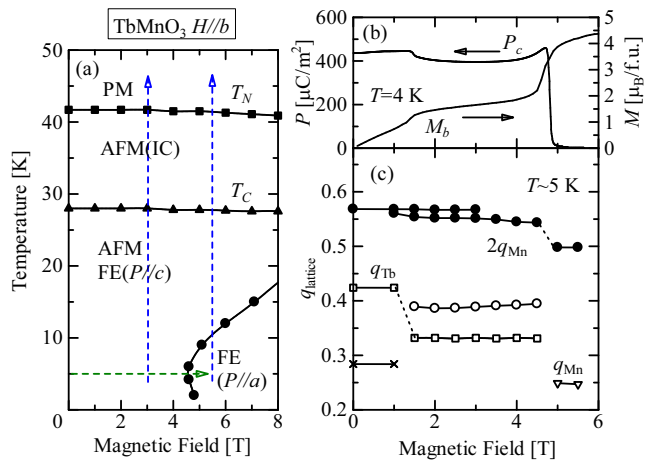}
\caption{
(Color online)
(a) Magnetic and electric phase diagram of TbMnO$_3$ with applied magnetic fields along the $b$ axis.  Squares, triangles, and circles indicate paramagnetic (PM)-to-incommensurate antiferromagnetic (AFM(IC)), AFM(IC)-to-{\boldmath $P$}$\parallel c$ ferroelectric (FE($P//c$)), and  FE($P//c$)-to-{\boldmath $P$}$\parallel a$ ferroelectric (FE($P//a$)) phase transitions, respectively.  
X-ray studies were performed in the condition along broken lines.  
(b) Polarization along the $c$ axis ($P_c$) and magnetization ($M_b$) at 4 K plotted against magnetic field applied parallel to the $b$ axis.  
(c) Lattice modulation vectors of TbMnO$_3$ at 5 K as a function of magnetic field. 
}
\label{tmo}
\end{figure}

A similar behavior is also observed in TbMnO$_3$ (Fig.~\ref{model}(a)).  
In a magnetic field of 3 T, the change in lattice modulation vector with temperature is almost the same as the previously reported results without a magnetic field.\cite{Kimura_PRB1}  
This is reasonable because a weaker magnetic field than 4.5 T has almost no effect on the successive phase transitions in TbMnO$_3$ as shown in Fig.~\ref{tmo}(a). 
In a magnetic field of 5.5 T, an intense superlattice peak at (0 4+1/2 1) as well as a much weaker one at (0 4+1/4 1) was observed at low temperatures corresponding to the $\makebox{\boldmath $P$}\parallel a$ phase.   
The observed commensurate lattice modulation should be ascribed to the magnetic modulation with $q_{\rm Mn}= 1/4$, as in the case of GdMnO$_3$.  
The Mn-spin modulation with $q_{\rm Mn}=1/4$ also coincides with the $\makebox{\boldmath $P$}\parallel a$ phase in TbMnO$_3$.  
With increasing temperature, another peak at $k\sim 4.54$ become discernible at around 10 K and coexist with the commensurate reflections with changing the relative intensities.   
Above 13 K, the commensurate peaks disappear.  
This behavior corresponds well to the polarization flop phenomenon from $\makebox{\boldmath $P$}\parallel a$ to $\makebox{\boldmath $P$}\parallel c$.\cite{Kimura_nature}  


The ferroelectric phase transition induced by application of a magnetic field can be understood in terms of symmetry.\cite{Goodenough}  
Orthorhombically distorted perovskites $R$MnO$_3$ with $Pbnm$ space group have inversion centers at Mn sites.  
The spin arrangement of the canted ferromagnetic phase of GdMnO$_3$ is of the so-called $A$-type with a modulation vector of Mn spin moments of (001).\cite{Popov,Noda,Loidl}  
Every Mn spin moment is approximately directed to $\pm b$ as schematically drawn in Fig.~\ref{model}(b), and slightly canted to the $c$ axis due to Dzialoshinski-Moriya interaction.\cite{DM}   
This type of spin arrangement keeps the inversion operation with respect to each Mn site, which agrees with the paraelectric property of this phase.\cite{Goto}

One of the most probable spin arrangements in the $H_b$-induced $\makebox{\boldmath $P$}\parallel a$ phase is of canted antiferromagnetic type as shown in Fig.~\ref{model}(c).   Here the compound is noncentrosymmetric because every inversion center vanishes.  
The direction of {\boldmath $P$} is also predictable, because {\boldmath $P$} should be parallel to all the mirror planes and glide planes.  
There is a symmetry operation $n'$, which is a composite of time reversal and a $n$-glide reflection with the plane shown by a dotted-broken line. 
A plane bisecting two neighboring MnO$_2$ sheets normal to the $c$ axis allows a $b$-glide reflection.  
An original $b$-glide plane normal to the $a$ axis shown by a broken line in Fig.~\ref{model}(b), in contrast, vanishes in the antiferromagnetic phase.  
A composite symmetry operation $b'$ of the $b$-glide reflection and time reversal is also broken.
Therefore, {\boldmath $P$} in GdMnO$_3$ with an antiferromagnetic spin modulation of (0 1/4 1) should be parallel to the $a$ axis, which completely coincides with the previous observation.\cite{Kimura_PRB2}  

The essence of the glide-symmetry breaking is the value of modulation vector.  
The breaking of $b$- or $b'$-glide reflection normal to the $a$ axis does not depend on the details of the spin alignment but only on the magnetic unit-cell dimension of $a\times 4b \times c$.\cite{reason}  
There remain many other possibilities for Mn spin configuration, because only the spin modulation vector was determined in the present study.  For example, Ising-type antiferromagnetic alignment is the most simple model.  Helical spin order is also preferable in spin-frustrated systems.  
For the latter case, when the spin moments rotate within the $ab$ plane with the propagation vector along the $b$ axis as shown in Fig.~\ref{model}(d), the {\boldmath $P$} along the $a$ axis as observed is the direct consequence of Dzialoshinski-Moriya interaction.\cite{DM,Katsura}


\begin{figure}
\includegraphics*[width=7cm]{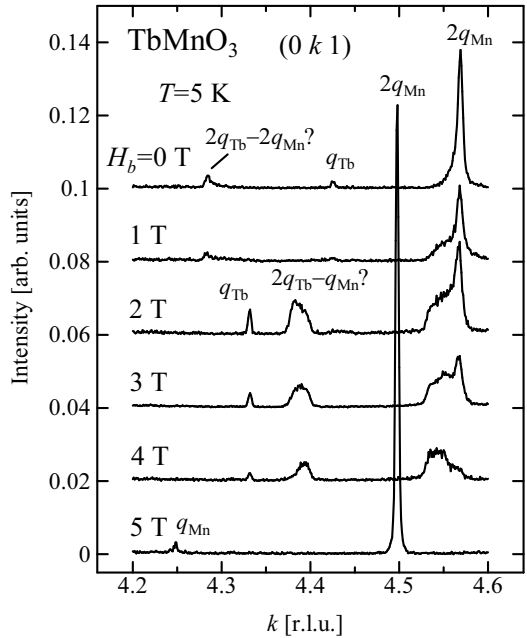}
\caption{
Change in synchrotron x-ray diffraction profile along the (0 $k$ 1) line in TbMnO$_3$ with increasing magnetic field at 5 K.  Peak assignments are given in terms of magnetic modulation vectors, $q_{\rm Mn}$ and $q_{\rm Tb}$.
}
\label{tmo_h}
\end{figure}

The zero-field ground state in TbMnO$_3$ is a ferroelectric phase with $\makebox{\boldmath $P$} \parallel c$ as shown in Fig.~\ref{tmo}(a).  
Application of magnetic field at low temperatures flop the {\boldmath $P$} vector from the $c$ to the $a$ direction.\cite{Kimura_nature}  
Change in superlattice peaks with the {\boldmath $P$}-flop phenomenon is shown in Fig.~\ref{tmo_h}.  
For the $H_b=0$ case, three peaks are discernible at $k=4.284$, 4.424, and 4.569.  Kajimoto et al.\ reported that fundamental modulation vectors for Mn and Tb moments are $q_{\rm Mn}=0.28$ and $q_{\rm Tb}=0.42$, respectively.\cite{Kajimoto}  
The lattice modulation at $k=4.569$ should be assigned to a conventional magnetostriction with {\boldmath $q_\ell$}=(0, $4+2q_{\rm Mn}$, 1).  
With increasing $H_b$ to 2 T, two peaks at $k=4.284$ and 4.424 disappear and the remaining peak at $k=4.569$ becomes broad.  
The peak at $k=4.284$ in weaker fields is hence not likely due to (0, $4+q_{\rm Mn}$, 1), but may be ascribed to some interference between two kinds of exchange striction related to Mn and Tb moments, such as (0, $4+2q_{\rm Tb}-2q_{\rm Mn}$, 1).  
Simultaneously one sharp peak and one broad peak appear at $k=4.333$ and at $k\sim 4.39$, respectively.  The former can be ascribed to the Tb-moment modulation with $q_{\rm Tb}=1/3$.  Since the latter new peak is about half as broad as the $q=2q_{\rm Mn}\sim 0.56$ peak and it suddenly appears with the change in Tb-moment modulation vector, we can assign it to lattice modulation with $2q_{\rm Tb}-q_{\rm Mn}$.  
The X-ray profile becomes quite simple above 5 T.  
One can observe two sharp peaks at commensurate positions of $k=4.50$ and 4.25.  The strong $q_\ell=1/2$ peak and weak $q_\ell=1/4$ peak should be ascribed to the exchange striction with $2q_{\rm Mn}$ and $q_{\rm Mn}$, respectively.  
In Fig.~\ref{tmo}(c), the lattice modulation vector $q_\ell$ is plotted as a function of $H_b$.  The above-mentioned discontinuous changes correspond well to magnetic anomalies around 2 T and 5 T in the magnetization curve.  
The metamagnetic anomaly of the magnetization curve at 5 T corresponds to the $H_b$ induced {\boldmath $P$} flop as well as to the commensurate locking to $q_{\rm Mn}=1/4$.  
The magnetic anomaly at 2 T is accompanied by a change in modulation vector of Tb spin moments (open squares).  
This clearly shows that Mn spin modulation is also affected by the change of $q_{\rm Tb}$ value, perhaps because the simple ratio of $q_{\rm Mn}$ to  $q_{\rm Tb}$ is violated.


The mechanism of ferroelectricity with $\makebox{\boldmath $P$} \parallel c$ in zero or weak fields seems to be more complicated than that of the $\makebox{\boldmath $P$} \parallel a$ phase.  
Locking of the modulation vector at the ferroelectric Curie temperature in TbMnO$_3$ and DyMnO$_3$ was suggested as the key to the $\makebox{\boldmath $P$} \parallel c$ ferroelectricity.\cite{Kimura_nature,Goto,Kimura_PRB2}  
This situation bears some analogy to improper ferroelectricity caused by a commensurate-incommensurate transition.\cite{improper}  
However, the present x-ray study in high magnetic fields seems to contradict such a scenario.  
The magnetic-field-induced {\boldmath $P$} flop in TbMnO$_3$ at 5 K occurs around 5 T, where the second anomaly is observed in the magnetization curve in Fig.~\ref{tmo}(b).  
Polarization along the $c$ axis subsists above 2 T up to 5 T, where the Mn-spin modulation vector becomes {\sl unlocked} with the change of $H_b$.  
Furthermore, the $q_{\rm Mn}$-value in the $\makebox{\boldmath $P$}\parallel c$ phase in 5.5 T is temperature-dependent down to 10 K as shown in Fig.~\ref{model}(a).  
It is also noteworthy that antiferromagnetic alignment of Ising spins along $\pm b$ at Mn sites with (0 $q_{\rm Mn}$ 1) does not break mirror reflections normal to the $c$ axis bisecting two adjacent MnO$_2$ planes, and hence it alone cannot produce the {\boldmath $P$} along the $c$ axis.  
A different explanation from the commensurate-incommensurate or locking mechanism has been proposed.\cite{Katsura}
More complicated spin configurations like helical ordering might produce {\boldmath $P$} along the $c$ axis.  In fact, a neutron study has reported that the magnetic structure in the $\makebox{\boldmath $P$} \parallel c$ phase is not of a simple sinusoidal type nor of an Ising type.\cite{Kajimoto}
A full spin-structure determination would be necessary to pin down the origin of magnetic ferroelectricity.

In summary, the magnetic-field-induced ferroelectric polarization along the $a$ axis observed both for GdMnO$_3$ and TbMnO$_3$ originates from the commensurate modulation of Mn spin moments with a modulation vector of (0 1/4 1), which is in good accordance with a group theory analysis.  
On the other hand, the $c$-polarized ferroelectricity of TbMnO$_3$ in zero or weak magnetic fields cannot be attributed to a locking of the spin modulation of Mn moments but perhaps to a noncollinear spin structure like helical order. 

We appreciate fruitful discussions with T. Kimura, S. Ishihara, and N. Nagaosa.  
This work was partly supported by Grant-In-Aids for Scientific Research from Ministry of Education, Culture, Sports, Science and Technology, Japan.  


\begin{thebibliography}{}
\bibitem{reason} In the antiferromagnetic state with {\boldmath $q$}=(0 1/4 1), $b$-glide reflection is defined as a composite operation of mirror reflection and translation of $2b$ (half of the magnetic period) but not $b/2$.
\end{thebibliography}
\end{document}